\documentclass[conference]{IEEEtran}
\IEEEoverridecommandlockouts
\usepackage{cite}
\usepackage{amsmath,amssymb,amsfonts}
\usepackage{accents}
\usepackage{comment}
\usepackage{bm}
\usepackage{bbm}
\usepackage{graphicx}
\usepackage{textcomp}
\usepackage{xcolor}
\usepackage{algorithmicx}
\usepackage{algorithm}
\usepackage{flushend}
\usepackage{subcaption}
\usepackage{mathtools}
\usepackage{amsthm}
\usepackage{verbatim}
\DeclareMathOperator*{\argmax}{arg\,max}
\DeclareMathOperator*{\argmin}{arg\,min}

\def\BibTeX{{\rm B\kern-.05em{\sc i\kern-.025em b}\kern-.08em
    T\kern-.1667em\lower.7ex\hbox{E}\kern-.125emX}}
\begin{document}

\title{Sub-Resolution mmWave FMCW Radar-based Touch Localization using Deep Learning\\
\thanks{R. M. Rao, K. A. Manjunatha, M. Hsu, and R. Kumar are with Amazon Lab126, Sunnyvale, CA, USA, 94089 (email:\{raghmrao, koushiam\}@amazon.com, mhsu@lab126.com, rrohk@amazon.com).}
\thanks{A. Kachroo is with Amazon Web Services (AWS), Santa Clara, CA, 95054 (email: amkachro@amazon.com). }
}

\author{\IEEEauthorblockN{Raghunandan M. Rao, Amit Kachroo,  Koushik A. Manjunatha, Morris Hsu, Rohit Kumar}
}

\maketitle

\begin{abstract}
Touchscreen-based interaction on display devices are ubiquitous nowadays. However, capacitive touch screens, the core technology that enables its widespread use, are prohibitively expensive to be used in large displays because the cost increases proportionally with the screen area. In this paper, we propose a millimeter wave (mmWave) radar-based solution to achieve sub-resolution error performance using a network of four mmWave radar sensors. Unfortunately, achieving this is non-trivial due to inherent range resolution limitations of mmWave radars, since the target (human hand, finger etc.) is \textit{`distributed'} in space. We overcome this using a deep learning-based approach, wherein we train a deep convolutional neural network (CNN) on range-FFT (range vs power profile)-based features against ground truth (GT) positions obtained using a capacitive touch screen. To emulate the clutter characteristics encountered in radar-based positioning of human fingers, we use a metallic finger mounted on a metallic robot arm as the target. Using this setup, we demonstrate sub-resolution position error performance. Compared to conventional signal processing (CSP)-based approaches, we achieve a $2-3\times$ reduction in positioning error using the CNN. Furthermore, we observe that the inference time performance and CNN model size support real-time integration of our approach on general purpose processor-based computing platforms.    
\end{abstract}

\vspace{-6pt}
\section{Introduction}
Modern displays use capacitive touchscreens for enabling touch-based interaction with the device, wherein touch localization is performed by processing the changes in electrical properties of the touchscreen layers across the display \cite{Lin_CapTouch_Stylus_JDT_2013}. In general, the touchscreen cost scales linearly with the area of the display covered by the touchscreen. As a result, it becomes prohibitively expensive to use capacitive touchscreens in large displays (for instance, display size $> 15$ inch). Furthermore, since the size of interactive elements (icons, sliders, buttons, etc.) tend to be large on a large screen, the positioning error requirement can often be relaxed from the typical mm-level accuracy to a few cm, without significantly impacting the user experience (UX). This work is focused on enabling accurate touch positioning in the latter regime.
\subsection{Related Work}
To reduce the cost while providing accurate positioning performance, there is significant interest in using alternative technologies such as Ultrasound \cite{yun_strata_mobisys_2017}, WiFi \cite{wu_fingerdraw_ACM_IMWUT_2020}, Radio Frequency \cite{wang_rf_idraw_acm_sigcomm_2014, shangguan_Polarize_CoNEXT_2016}, mmWave \cite{wei_mtrack_mobicom_2015, xiao_milliback_acm_2019} and Ultrawideband (UWB) \cite{cao_itracku_acm_2021, Nermine_DL_AirWrite_IEEE_SJ_2022, Faheem_in_air_write_IEEE_Acc_2020} radar. Yun \textit{et al.} \cite{yun_strata_mobisys_2017} designed a device-free system using ultrasound signals to track human finger motion. Their algorithm is based on processing the channel impulse response (CIR) to estimate the absolute distance and the distance displacement using multiple CIRs, resulting in a \textit{median tracking error} $\delta r_{\rm track, 50}=1$ cm. Wu \textit{et al.} \cite{wu_fingerdraw_ACM_IMWUT_2020} proposed a sub-wavelength finger motion tracking system using one WiFi transmitter and two WiFi receivers. Their approach used a channel quotient-based feature to detect minute changes in the channel state information (CSI) due to finger movement, resulting in $\delta r_{\rm track, 90} = 6$ cm. Wang \textit{et al.} \cite{wang_rf_idraw_acm_sigcomm_2014} propose a Radio Frequency Identification (RFID) sensor worn on the finger to demonstrate precise tracking of air handwriting gestures. The authors demonstrated tracking errors of $\delta r_{\rm track, 90} = 9.7$ cm and $10.5$ cm in Line-of-Sight (LoS) and Non-LoS (NLoS) conditions respectively. Shangguan \cite{shangguan_Polarize_CoNEXT_2016} proposed an air handwriting system based on two differentially polarized antennas to track the orientation and position of an RFID-tagged pen, achieving a \textit{$90^\text{th}$ percentile position error} ($\delta r_{\rm pos, 90}$) of $11$ cm. Xiao \textit{et al.} \cite{xiao_milliback_acm_2019} proposed an RF backscattering-based system to track handwriting traces performed using a stylus in which a RFID tag is embedded. The authors demonstrated $\delta r_{\rm track, 50} = 0.49$ cm at a writing speed of $30$ cm/s. Wei \cite{wei_mtrack_mobicom_2015} designed a $60$ GHz radio-based passive tracking system to position different writing objects such as pen/pencil/marker to obtain $\delta r_{\rm pos, 90} = 0.8/5/15$ cm respectively. Their approach relies on passive backscattering of a single carrier $60$ GHz signal, using which the initial location is acquired. The low tracking error is obtained by tracking its phase over time.  In \cite{cao_itracku_acm_2021}, the authors propose an Inertial Measurement Unit (IMU)-UWB radar fusion-based tracking approach to implement a stylus-aided handwriting use-case that achieves $\delta r_{\rm pos, 50} = 0.49$ cm. However, in the absence of the IMU, the authors report $\delta_{\rm pos, 90} = 6$ cm. 

Even though the works \cite{yun_strata_mobisys_2017, wu_fingerdraw_ACM_IMWUT_2020, wang_rf_idraw_acm_sigcomm_2014, xiao_milliback_acm_2019} report a low tracking error, the position error is high. While this trade-off is acceptable in finger tracking applications such as handwriting recognition, it is unacceptable for on-screen interaction where the \textit{performance is dictated by the position error, not by the accuracy of the reconstructed trajectory}. On the other hand, works such as \cite{cao_itracku_acm_2021} that achieve cm-level position accuracy necessitates the use of additional IMU sensors, that adds system/computational complexity, and friction to the UX.
\vspace{-5pt}
\subsection{Contributions}
In this work, we bridge this gap by proposing a mmWave radar sensor network-based positioning framework that uses a Deep Convolutional Neural Network (CNN) to achieve sub-resolution position accuracy. We build a robot-based testbed for characterizing positioning performance, in which we use a robot-mounted metal finger as the distributed target. We collect data for multiple runs to capture the metal finger's signature for each radar sensor at different locations of the screen, as well as the corresponding ground truth position using a capacitive touchscreen. The conventional signal processing (CSP)-based approach that uses range calibration coupled with the nonlinear least squares (NLS) algorithm  \cite{buehrer2019handbook} results in $\delta r_{\rm pos, 90} = 3.7$ cm. On the other hand, inspired by the LeNet model \cite{LeNet5_YLC_Proc_IEEE_1998}, we design a CNN that outperforms the CSP-based approach and achieves sub-resolution accuracy, with $\delta r_{\rm pos, 90} = 1.6$ cm. Finally, we demonstrate that the small model size and CNN inference time makes real-time implementation feasible on general purpose processor-based computing platforms. 

\begin{figure*}[!t]
	\centering
	\begin{subfigure}[t]{0.52\textwidth}
		\centering
		\includegraphics[width=\textwidth]{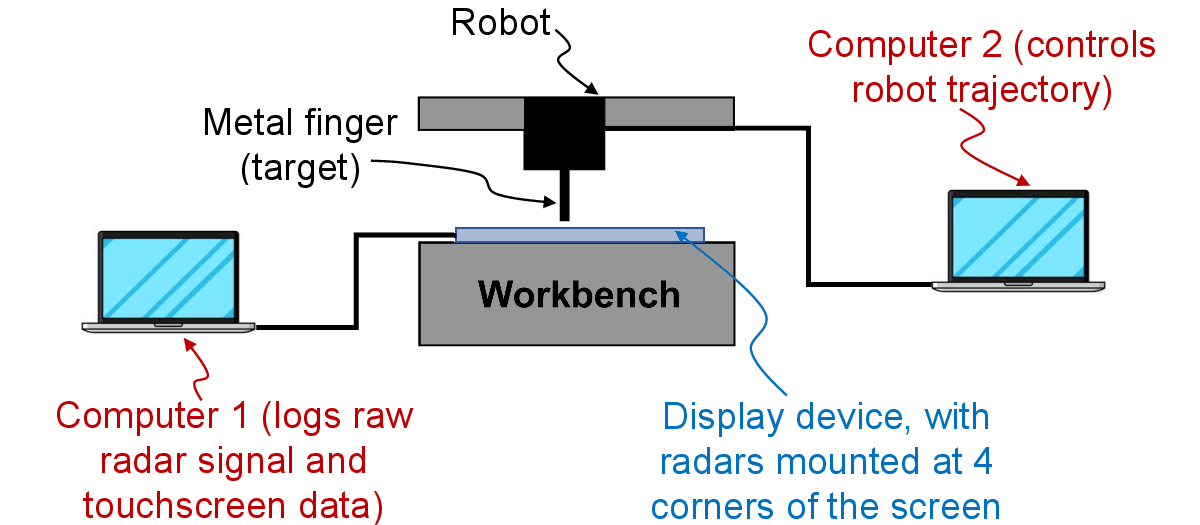}\\
		[-1ex]
		\caption{}
		\label{subfig:robot_setup}
	\end{subfigure}
	\hfill
	\begin{subfigure}[t]{0.46\textwidth}
		\raggedleft
		\includegraphics[width=\textwidth]{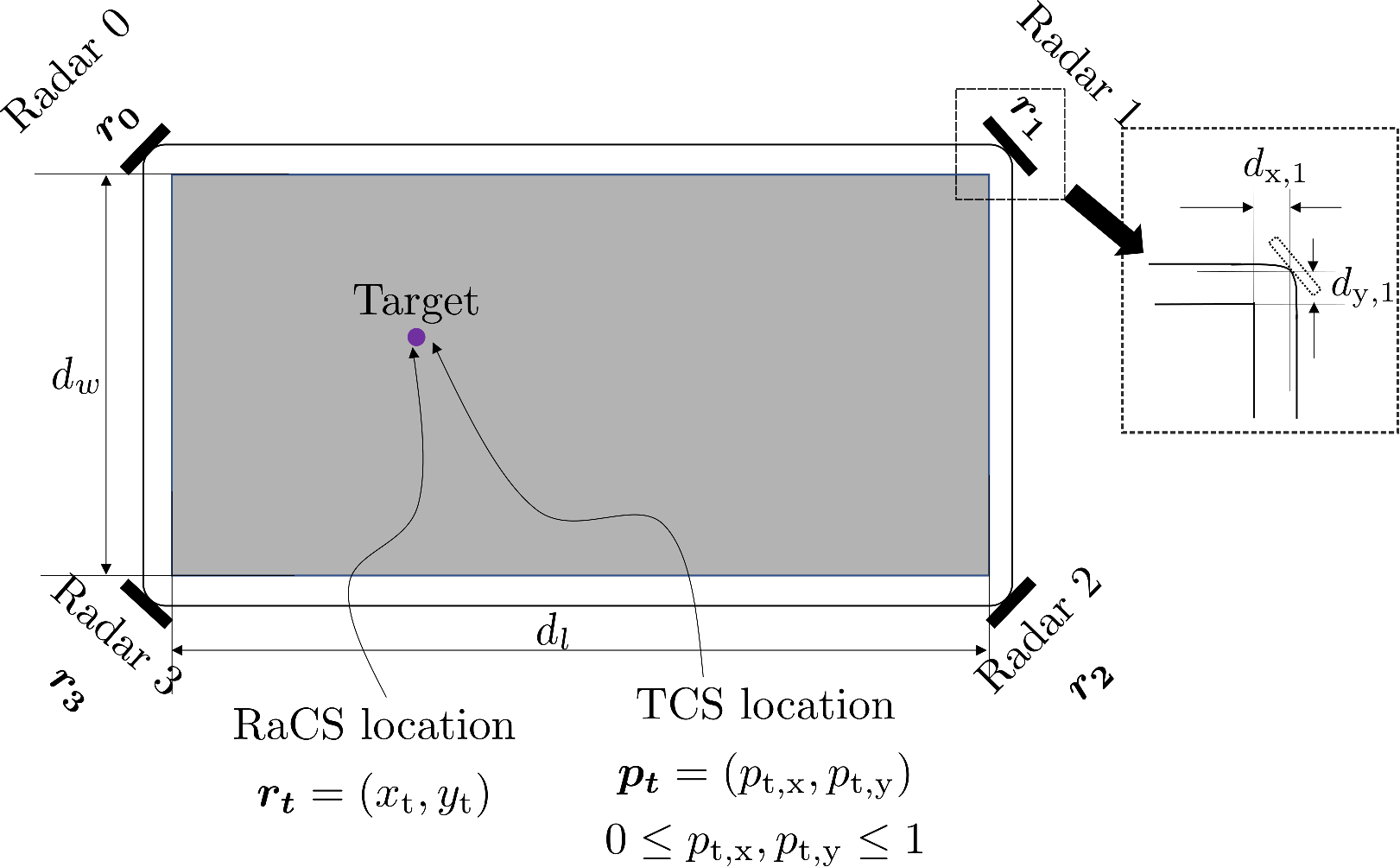}\\
		[-1ex]
		\caption{}
		\label{subfig:schematic_of_setup}
	\end{subfigure}
	\caption{Schematic of (a) the robot-based data collection setup, and (b) mmWave radar network-based positioning setup, and description of the \textbf{Ra}dar \textbf{C}oordinate \textbf{S}ystem (RaCS) and \textbf{T}ouchscreen \textbf{C}oordinate \textbf{S}ystem (TCS).}
	\label{fig:radar_touch_testbed}
\end{figure*}

\begin{figure}[!ht]
	\centering
	\includegraphics[width=0.8\linewidth]{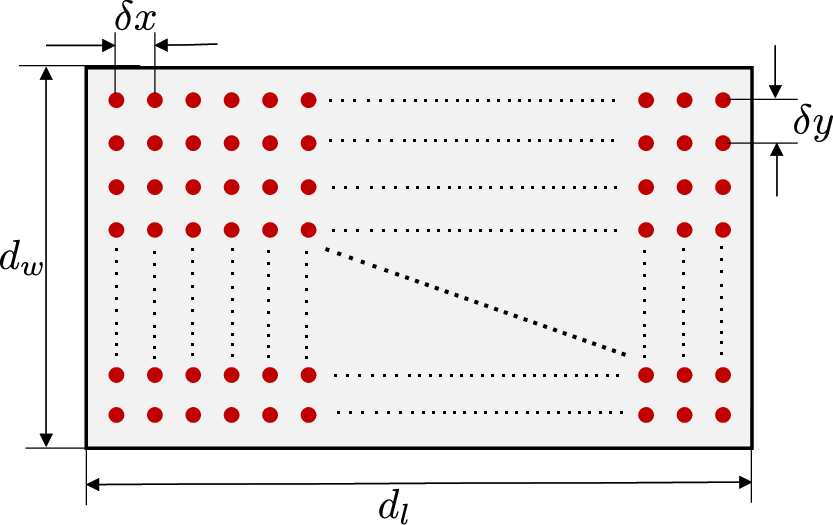}\\
	[-1 ex]
	\caption{Schematic of the touch locations on the display. In our setup, $d_{l} = 34.3$ cm, $d_w = 17.8$ cm, and $\delta x = \delta y = 1$ cm.}
	\label{fig:touch_points_schematic}
\end{figure}

\section{System Design}
\subsection{Working Principle}
In contemporary consumer electronic devices, touch localization on a capacitive touchscreen displays rely on changes in the electrical properties of carefully arranged material layers when a finger touches the screen. In essence, the location is estimated by determining the `touch cell' where there is maximum variation in the capacitance \cite{Shen_CAuth_IEEE_TKDE_2023}. In contrast, contactless methods can also be used if accurate distance \cite {Yan_Tiberius_AESM_2013, Rao_Bey_Conv_Hull_TVT_2021, Rao_RNDOP_TVT_2024} and/or angle information \cite{Badriasl_3D_AoA_IEEE_TAES_2014} of the finger is known relative to sensors \textit{with known positions}. In this work, we design a low-cost touch positioning system that uses multiple Frequency Modulated Continuous Wave (FMCW) mmWave radar sensors to locate the ``finger'' (target) on the screen. FMCW mmWave radar sensors are attractive for short range sensing applications because they can be operated at low-power and manufactured with low cost. This is in part due to the low bandwidth ($\sim 1$ MHz) of the baseband signal processing chain, despite the large sweep bandwidth ($> 1$ GHz) \cite{Santra_Hazra_DL_Radars_2020}.  

However, the main factor that limits accurate finger localization is the limited range resolution ($\Delta r_{\rm res}$) of the radar, given by $\Delta r_{\rm res} = \frac{c}{2 f_{\rm BW}}$, where $c$ is the free-space velocity of light, and $f_{\rm BW}$ the chirp bandwidth of the FMCW radar \cite{Uysal_PC_FMCW_IEEE_TVT_2020}. For example, the radar will not be able to distinguish objects that are closer than $3$ cm (in the radial direction) for $f_{\rm BW} = 5$ GHz. As a result, the finger and the rest of the hand often appear as a single target to each radar sensor, thus resulting in a range error that is dictated by $\Delta r_{\rm res}$.

\subsection{Experimental Testbed Setup}
The main focus of this work is to evaluate the position error of a mmWave radar-based touch solution. To undertake this, we built a testbed whose schematic is shown in Fig. \ref{subfig:robot_setup}. The setup is based on a 15.6 inch display, on which $N_{\rm rad} = 4$ radar sensors are placed at the corners of the screen using 3D printed fixtures, as shown in Fig. \ref{subfig:schematic_of_setup}. The display is equipped with a capacitive touchscreen, which is interfaced with a dedicated computer to obtain the ground truth (position and time) for each touch event. A metal finger is used as the target which is to be localized by the radar sensor network. The metal finger is mounted on a programmable robot to achieve precise control over its trajectory during the data collection session, as shown in Fig. \ref{subfig:robot_setup}. The robot is controlled by another dedicated computer, and is programmed to touch the display on a grid of points, as shown in Fig. \ref{fig:touch_points_schematic}. The spacing between each point on the grid is approximately $1 \text{ cm}$ along both vertical and horizontal directions. It is important to note that localization performance in this setup is typically limited by $\Delta r_{\rm res}$, since the metal finger (target, analogous to the human finger) and the metallic robotic arm (analogous to the rest of the hand) on which it is mounted \textit{will appear as a single target} to the radar. The radar configuration used is shown in Table \ref{table:radar_sensor_params}. It is worthwhile to note that this waveform is compliant with the latest FCC final rule \cite{FCC_Empows_60GHz_Rad}. 

\begin{table}[t]
	\centering		
	\caption{Radar Sensor Parameters}
	\label{table:radar_sensor_params}
	\begin{tabular}{|l|l|} 
		\hline 
		\textbf{Parameter} & \textbf{Value} \\
		\hline
		Waveform type & FMCW \\
		\hline
		Chirp Bandwidth ($f_{\rm BW}$) & $4.874$ GHz \\
		\hline 
		Range Resolution ($\Delta r$) & $3.075$ cm \\ 
		\hline 
		Frame Rate ($f_{\rm r}$) & $120$ Hz \\
		\hline 
		Number of IF samples/chirp ($N_{\rm IF}$) & 64 \\
		\hline 
		Number of chirps/frame ($N_{\rm ch}$) & 8 \\
		\hline
		Number of RX antennas/sensor ($N_{\rm rx}$) & 3 \\
		\hline 
		Number of radar sensors ($N_{\rm rad}$) & 4 \\
		\hline  
	\end{tabular}
\end{table}

\begin{figure*}[!ht]
    \centering
    \includegraphics[width=0.9\textwidth]{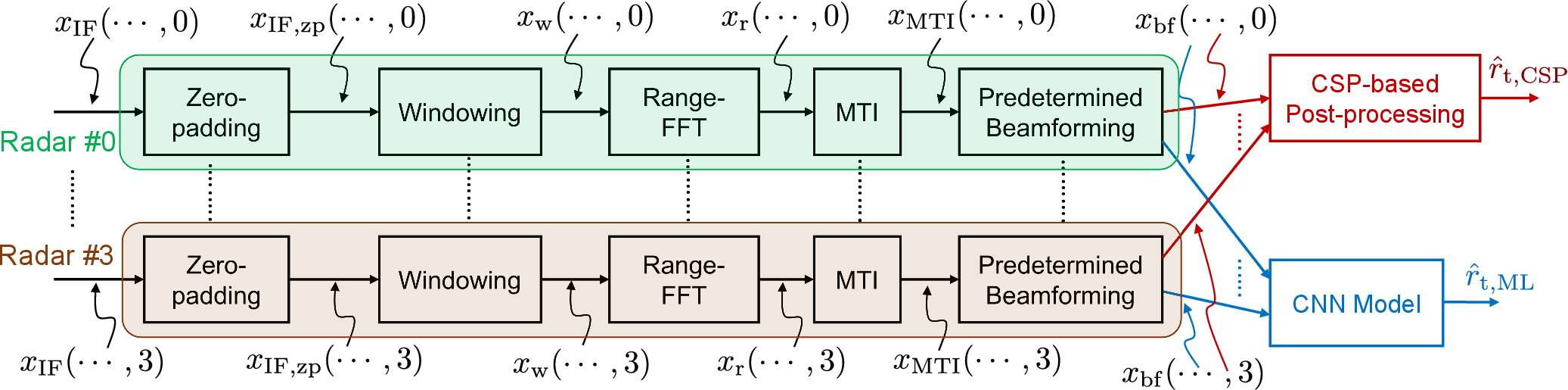}
    \caption{Flowchart of the radar signal processing pipeline implemented on each radar sensor.}
    \label{fig4:sig_proc_chain}
\end{figure*}
\vspace{-5pt}
\subsection{Radar Signal Pre-Processing}
The signal processing pipeline for generating the feature is shown in Fig. \ref{fig4:sig_proc_chain}. Let $f, s, r, c, j$ and $i$ denote the frame index, IF sample index, range bin, chirp index, RX antenna index, and the sensor index respectively. Each radar sensor transmits the FMCW waveform with parameters shown in Table \ref{table:radar_sensor_params}. For each frame $f$, the received waveform is then down-converted to get the  intermediate frequency (IF) signal $x_{\rm IF}(f,s,c,j,i)$, that corresponds to the radar return. From this, the range information corresponding to all scattering objects in the radar's field of view (FoV) is obtained by computing the beamformed \textit{`range-FFT'} $x_{\rm r}(f,r,c,j)$ using the following sequence of steps.
\begin{align}
	\label{eq:zero_padding}
	&x_{\rm IF,zp}(f,s,c,j,i) = \begin{cases}
		x_{\rm IF}(f,s,c,j,i) \text{ for } s < N_{\rm IF} \\
		0 \text{ for } N_{\rm IF} \leq s \leq N_{\rm os} N_{\rm IF} - 1
	\end{cases}, \\
	\label{eq:windowing}
	&x_{\rm w}(f,s,c,j,i) = x_{\rm IF,zp}(f,s,c,j,i) w_{\rm IF}(s), \\
	\label{eq:range_fft}
	&x_{\rm r} (f,r,c,j,i) = \sum\limits_{s = 0}^{N_{\rm os} N_{\rm IF} - 1} x_{\rm w} (f,s,c,j,i) e^{\frac{-j2 \pi s r}{N_{\rm os} N_{\rm IF}}}, 
\end{align}   
for $r=0,1,\cdots,\Big(\frac{N_{\rm os} N_{\rm IF}}{2} - 1 \Big)$. Here, zero-padding is performed in (\ref{eq:zero_padding}) to shrink the effective range-bin width to $\Delta r_{\rm os} = \Delta r/N_{\rm os}$\footnote{Note that while this shrinks the range bin width, the range resolution (i.e. minimum radial distance between two targets such that they appear as two distinct targets) is unchanged. Oversampling in the range domain minimizes the contribution of range quantization error in the system.}, where $N_{\rm os}=8$ is the oversampling factor. The zero-padded signal is then used to compute the range-FFT in (\ref{eq:range_fft}) after a windowing operation $w_{\rm IF}(\cdot)$. The purpose of the latter is to trade-off the range-FFT sidelobe level with the mainlobe width. It is worthwhile to note that the IF signal contains only the in-phase component and hence, is a real-valued signal. Thus, the range-FFT is symmetric about  $r = N_{\rm IF} N_{\rm os}/2$. Clutter removal is then used to eliminate the scattered returns from static objects using an IIR moving target indication (MTI) filter to get the post-MTI range-FFT signal $x_{\rm MTI}(f,r,c,j,i)$, given by
\begin{align}
	\label{eq:MTI_operation}
	x_{\rm c}(f,r,c,j,i) & = \beta x_{\rm r}(f,r,c,j,i) + (1 - \beta)  x_{\rm c}(f,r,c,j,i), \nonumber \\
	x_{\rm MTI}(f,r,c,j,i) & = x_{\rm r}(f,r,c,j,i) - 	x_{\rm c}(f,r,c,j,i),
\end{align}
where $0 < \beta < 1$ is the IIR filter response parameter and $x_{\rm c}(f,\cdots)$ is the clutter estimate for the $f^\text{th}$ frame. Finally, to keep the feature dimension manageable for real-time implementation, averaging across chirps and \textit{boresight beamforming} are performed to get the beamformed range-FFT feature $x_{\rm bf}(f,r,i)$ using\footnote{Since the feature is obtained through linear operations on the raw IF signal, averaging across chirps and boresight beamforming can equivalently be performed on the IF signal prior to range-fft as well. } 
\begin{align}
	\label{eq:avg_across_chirp_boresight_bf}
	x_{\rm bf}(f,r,i) = \frac{1}{N_{\rm rx} N_{\rm ch}}\sum_{j=0}^{N_{\rm rx} - 1} \sum_{c = 0}^{N_{\rm ch} - 1} x_{\rm MTI}(f,r,c,j,i). 
\end{align}
Note that for uniform linear/planar arrays, boresight beamforming is equivalent to signal averaging across RX antennas. 

\subsection{Ground Truth}
For each touch event, the capacitive touchscreen-based ground truth (GT) information is composed of the relative location $(p_{\rm t,x}, p_{\rm t, y})$ and touch timestamp ($t_{\rm GT}$), such that $0 \leq p_{\rm t,x} \leq 1$ and  $0 \leq p_{\rm t,y} \leq 1$. The relative locations are converted to locations in the radar coordinate system $\bm{r}_{\rm GT} = (x_{\rm t, GT}, y_{\rm t,GT})$ using knowledge of the reference radar location w.r.t. the touch area. We use the sign convention shown in Fig \ref{subfig:schematic_of_setup}, using which the GT coordinates are calculated using 
\begin{align}
	\label{eq:GT_coords_from_rel_coords}
	\bm{r}_{\rm t, GT} = (r_{\rm t,x} d_l + d_{\rm x,0}, -r_{\rm t,y} d_w - d_{\rm y,0}). 
 \end{align}

\begin{figure*}[ht]
	\centering
	\includegraphics[width=0.9\linewidth]{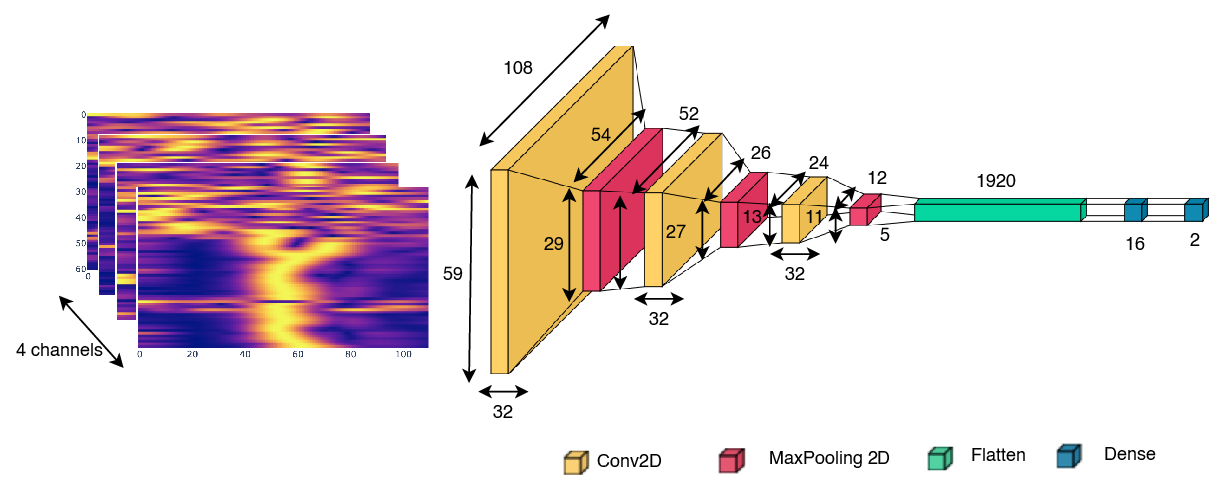}\\
	[-2ex]
	\caption{The four-channel input ($\bar{\bm{x}}_{\bm{bf}}(\mathcal{F}_i(n), \mathcal{R}) )$ consists of heatmaps from the four mmWave radars. The Deep Convolutional Neural Network consisting of three Convolution2D layers and Maxpooling2D and one Dense layer. The final dense layer outputs the 2D position estimate ($\hat{\bm{r}}_{\rm t, ML}$) of the touch event on the display.}
	\label{fig:dnn_architecture}
\end{figure*}
\vspace{-10pt}
\section{Conventional Signal Processing (CSP)-based Positioning}
To improve the accuracy of the conventional signal processing-based estimates, we use range estimates from the beamformed signal, as well as the per-RX signals. Firstly, the post-MTI range-FFT from each RX antenna is averaged across chirps using
\begin{align}
	\label{eq:per_RX_signals}
	x_{\rm MTI,c}(f,r,j,i) & = \frac{1}{N_{\rm ch}} \sum_{c = 0}^{N_{\rm ch} - 1} x_{\rm MTI}(f,r,c,j,i). 
\end{align}

Then, range estimates corresponding to the per-RX ($\hat{r}_{\rm ij}(f)$) as well as beamformed signals ($\hat{r}_{\rm bf,i}(f)$) are estimated using
\begin{align}
	\label{eq:range_estims_per_RX_bf}
	\hat{r}_{\rm ij}(f) & = \Delta r_{\rm os} \cdot  \underset{r}{\argmax}\ |x_{\rm MTI,c}(f,r,j,i)|^2, \\
	\hat{r}_{\rm bf,i}(f) & = \Delta r_{\rm os} \cdot \underset{r}{\argmax}\ |x_{\rm bf}(f,r,i)|^2.
\end{align}

To have reliable ranging performance in the presence of low SNR conditions and strong clutter regions (e.g. portion of the hand excluding the finger such as shoulder, torso, palm etc.), we invalidate the range estimate when there is no consensus among the different per-RX range estimates. The range estimate from the $i^\text{th}$ sensor is computed using
\begin{align}
	\label{eq:range_RX_consensus}
	\hat{r}_{\rm i}(f) = \begin{cases}
		\hat{r}_{\rm bf,i}(f) - r_{\rm cal, i} & \text{if } |\hat{r}_{\rm ij}(f) - \hat{r}_{\rm ik}(f)| \leq \Delta r_{\rm th} \\
		& \forall j \neq k  \\
		{\rm nan} & \text{ otherwise.}
	\end{cases},
\end{align}
where $\Delta r_{\rm th}$ is the range consensus tolerance, and $r_{\rm cal,i}$ is the range offset for the $i^{\rm th}$ sensor, which is obtained using a one-time calibration of the localization environment. Let $f_n$ be the radar frame index corresponding to the $n^\text{th}$ touch event. Then, the range estimates are averaged over a window of $N_w = 5$ frames, to mitigate the unavailability of range estimates, resulting in a range estimate $\bar{r}_{\rm i} (f_n) = \frac{1}{N_{\rm val}} \sum_{m = 0}^{N_w - 1} \hat{r}_{\rm i}(f_n - m) \mathbbm{1}[\hat{r}_{\rm i}(f_n - m) \neq {\rm nan}]$, where $N_{\rm val} = |\{m|\hat{r}_{\rm i}(f_n - m) \neq {\rm nan}\}|$.

Finally, the target's position estimate ($\hat{\bm{r}}_{\rm t, CSP}(f)$) is obtained by solving the nonlinear least squares (NLS) problem \cite{buehrer2019handbook}  

\begin{align}
	\label{eq:NLS} 
	\hat{\bm{r}}_{\rm t, CSP}(f_n) = \underset{\bm{r}}{\argmin}\ \sum_{i=0}^{N_{\rm s} - 1} (\| \bm{r_i} - \bm{r} \|_2 - \bar{r}_{\rm i}(f_n))^2. 
\end{align}

\begin{figure}[t]
	\centering
	\includegraphics[width=\linewidth]{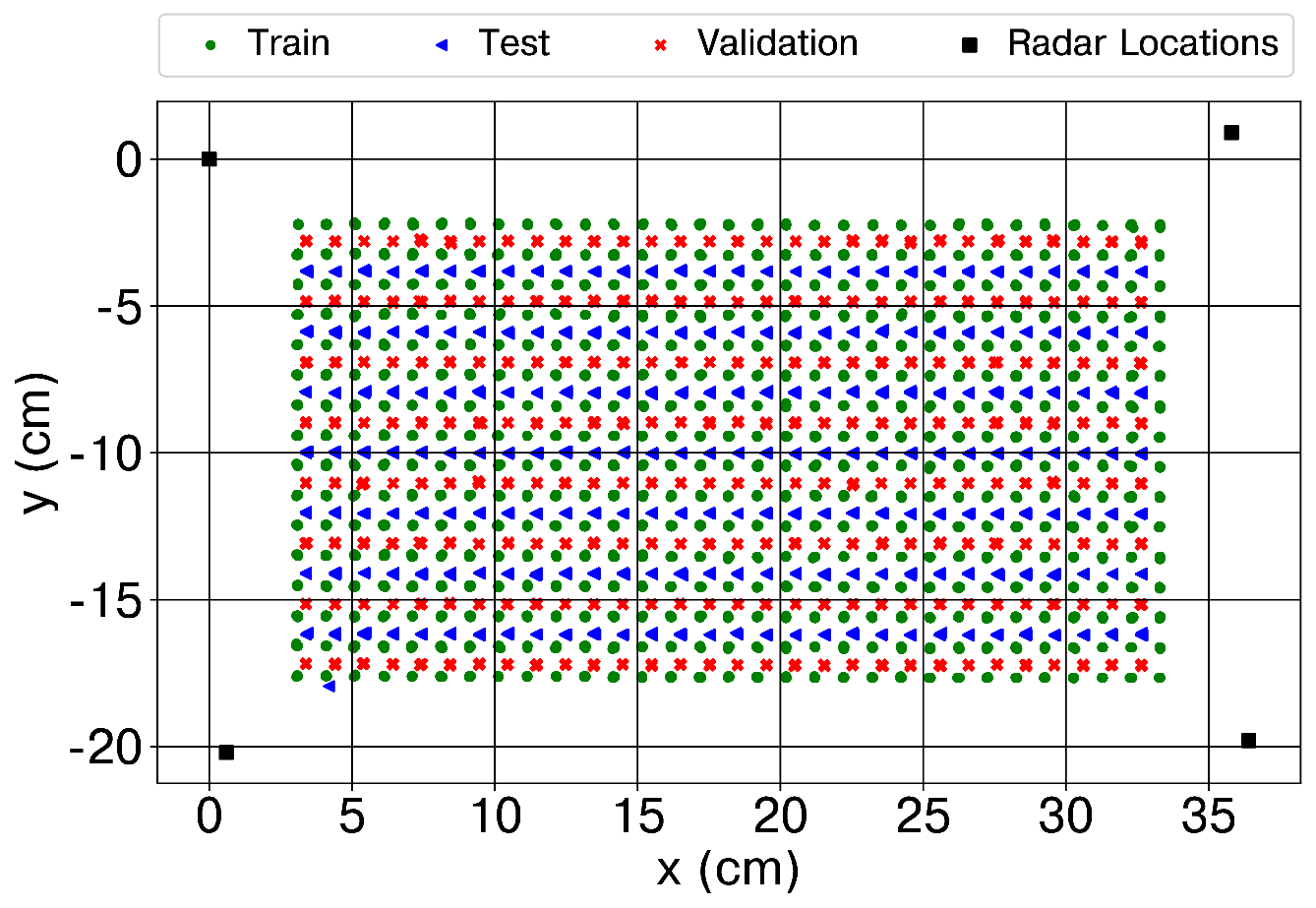}
	\caption{Partitioning of touch points into train/validation/test datasets, and their locations relative to the radar sensors. In our testbed, the radar locations are $\bm{r_0} = (0,0))$ cm, $\bm{r_1} = (35.8, 0.9)$ cm, $\bm{r_2} = (36.4, -19.8)$ cm, and $\bm{r_3} = (0.6,-20.2)$ cm.}
	\label{fig:val_and_test_points}
\end{figure}
\vspace{-5pt}
\section{Deep Neural Network-based Positioning}
\subsection{Feature Generation}
The datastream from each radar sensor and the capacitive touchscreen are collected independently without explicit synchronization. The relatively high sampling rate of the radar ($120$ Hz) and the touchscreen ($90$ Hz) w.r.t. the finger movement speed eliminates the need for explicit sensor synchronization. The radar frame indices corresponding to each touch event are found using the GT touch time $t_{\rm GT}$. Suppose the $i^\text{th}$ radar frame index corresponding to the touch event is $f_{\rm GT,i}$ For each touch event, the feature contains the beamformed range-FFT for all radar sensors for the frames $\mathcal{F}_i=\{f_{\rm GT,i}-25, f_{\rm GT,i}-24, \cdots, f_{\rm GT,i}+25 \}$ for the range bin indices $\mathcal{R} = \{0,1,\cdots, R_{\max}\}$, where $R_{\max}=\Big \lceil \frac{N_{\rm os} \sqrt{d^2_l + d^2_w}}{\Delta r} \Big \rceil$ is the maximum possible target distance on the display. In our setup, $R_{\rm max} = 110$. The feature for the $n^\text{th}$ touch event is the tensor $\bar{\bm{x}}_{\bm{bf}}(\mathcal{F}_i(n), \mathcal{R}) \in \mathbb{C}^{61 \times 110 \times 4}$, comprising of four heatmaps, each corresponding to one of the radar sensors. 

\subsection{Machine Learning Model Architecture}
The proposed CNN-based architecture used in this work is shown in Fig. \ref{fig:dnn_architecture}, which is similar to the LeNet-5 architecture \cite{LeNet5_YLC_Proc_IEEE_1998} except it uses 3 layers of convolution with 3 max-pooling layers rather than 3 layers of convolution with 2 average pooling in the original architecture. We use three cascaded 2D-Convolutional + 2D-Maxpooling layers with filter sizes of $32, 32,$ and $32$ respectively, each with a kernel size of $3 \times 3$ with ReLu activation. The maxpooling layer after convolution is used not only to reduce the feature size but also reduce over-fitting. This  improves the generalization and also reduces the memory requirements to host the model on the device. After the convolution and pooling operations, the output is flattened, followed by a dense layer with $16$ units and then the final $2$ dense unit to generate the 2D position coordinates $\hat{\bm{r}}_{\rm  \bm{t, ML}} = (\hat{x}_{\rm t, ML}, \hat{y}_{\rm t, ML})$.
\vspace{-5pt}
\section{Experimental Results}
\subsection{Data Collection}
The effective touch area on the 15.6 inch display is a rectangular area of length $d_{l} = 34.3$ cm and width $d_w = 17.8$ cm. In a single session, the data is collected across a point grid with an arbitrary offset from Radar 0 ($\bm{r}_{\bm{0}}$), such that consecutive touch points are separated by $1$ cm along both the axes, as shown in Fig. \ref{fig:touch_points_schematic}. Data is collected in two stages:
\begin{enumerate}
	\item \textit{Training Dataset:} In this stage, we collected data for 50 sessions. In each session, the robot touched the screen across a $31 \times 16$ grid. After accumulating data across all sessions, the training data has the dimension $(24799,61,110,4)$. 
	\item \textit{Validation and Test Datasets:} In this stage, we collected data for 15 sessions. In each session, the robot touched the screen across a $30 \times 15$ grid. The grid pattern was designed such that the validation/test touchpoints have a position offset of $\bm{\delta r'}= (0.5 \text{ cm}, 0.5 \text{ cm})$ relative to the training touchpoints. This offset is introduced to test the generalization performance of the machine learning model to unseen data. Finally, touchpoints in the odd/even rows are allocated to the validation/test datasets, as shown in Fig. \ref{fig:val_and_test_points}. Hence, the validation and dataset dimensions are $(3600, 61, 110, 4)$ and $(3150, 61, 110, 4)$ respectively. 	
\end{enumerate}

\subsection{Range Calibration}
For the conventional signal processing approach, the range calibration offset for each radar sensor ($r_{\rm cal, i}$) is estimated using the training dataset. For the $n^\text{th}$ touchpoint in the training set, the range estimate corresponding to each radar sensor is computed using (\ref{eq:range_RX_consensus}), and is compared to the GT distance $r_{\rm t, GT, i}(n) = \| \bm{r}_{\rm t, GT}(n) - \bm{r_i} \|_2$. Suppose the corresponding range error is $\Delta r_i(n) = r_{\rm t, GT, i}(n) - \bar{r}_{\rm i} (f_n)$. Then, the range calibration offset is estimated by computing the empirical average of the range errors for each sensor, i.e. $r_{\rm cal, i} = \frac{1}{N_{\rm train}} \sum_{n = 0}^{N_{\rm train} - 1} \Delta r_i(n)$, where $N_{\rm train} = 24799$. Note that this is the least-squares (LS) solution of the range-bias estimation problem under the model $\hat{r}(n) = r(n) + \epsilon$ for $n = 0,1,\cdots,N_{\rm train} - 1$, where $\epsilon$ is the range bias.  

\begin{figure}[t]
	\centering
	\includegraphics[width=0.9\linewidth]{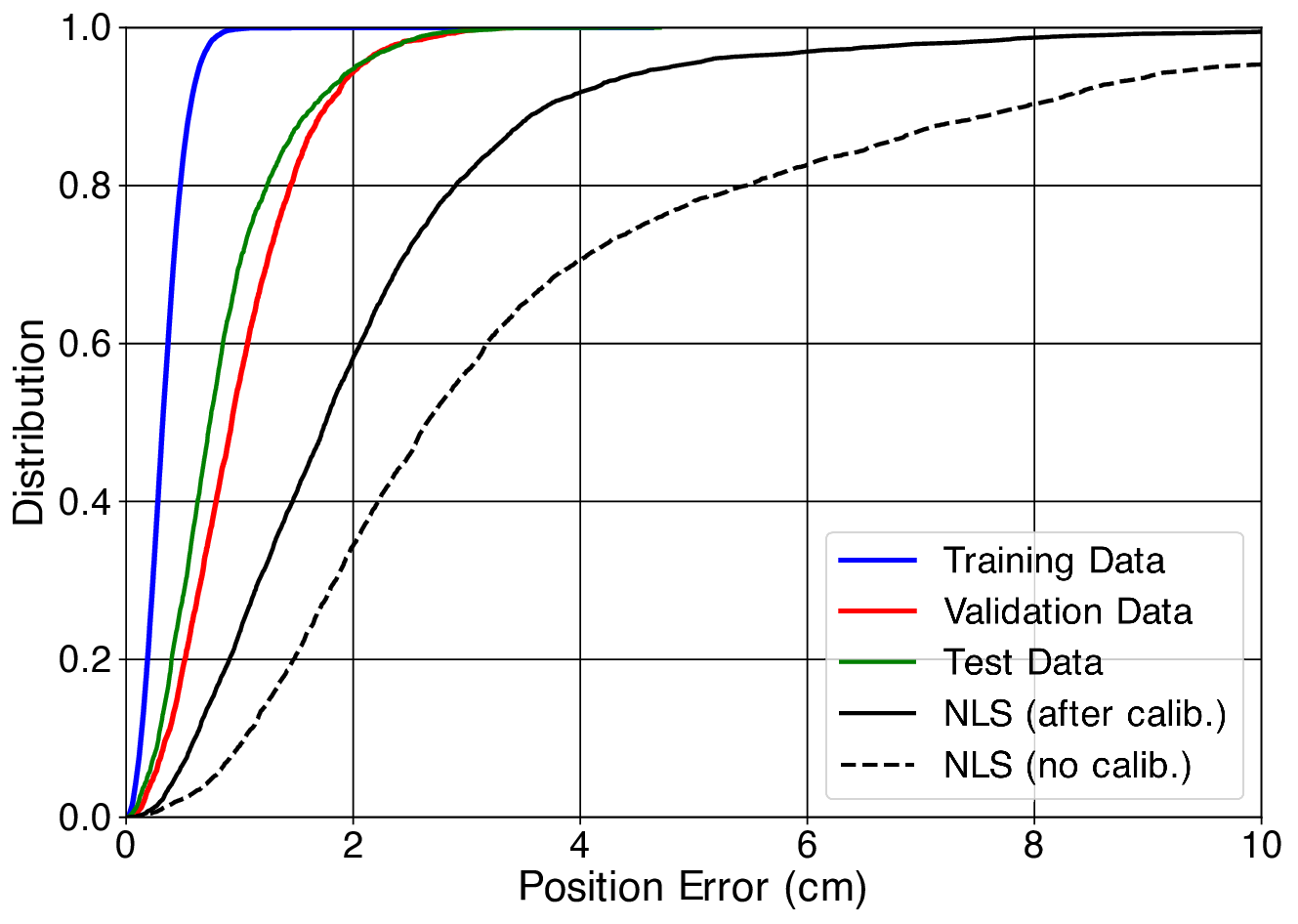}
	\caption{Distribution of position error for the Conventional Signal Processing and ML-based approaches.}
	\label{fig:pos_err_ML_vs_CSP_comparison}
\end{figure}

\begin{table}[t]
	\centering		
	\caption{Position Error Performance Comparison on the Test Dataset}
	\label{table:perf_comparison}
	\begin{tabular}{|l|l|l|} 
		\hline 
		\textbf{Performance Metric} & \textbf{CNN-based} & \textbf{CSP-based}\\
		\hline
		Median Pointwise RMSE &  $0.84$ cm & $1.92$ cm \\
		\hline 
		$90 \%$ile Pointwise RMSE &  $1.6$ cm & $3.85$ cm \\
		\hline 
		Median Error (all points) & $0.82$ cm & $1.75$ cm \\
		\hline 
		$90\%$ile Error (all points) & $1.62$ cm & $3.7$ cm \\
		\hline
	\end{tabular}
\end{table}

\begin{figure*}[!t]
	\centering
	\begin{subfigure}[t]{0.48\textwidth}
		\centering
		\includegraphics[width=\textwidth]{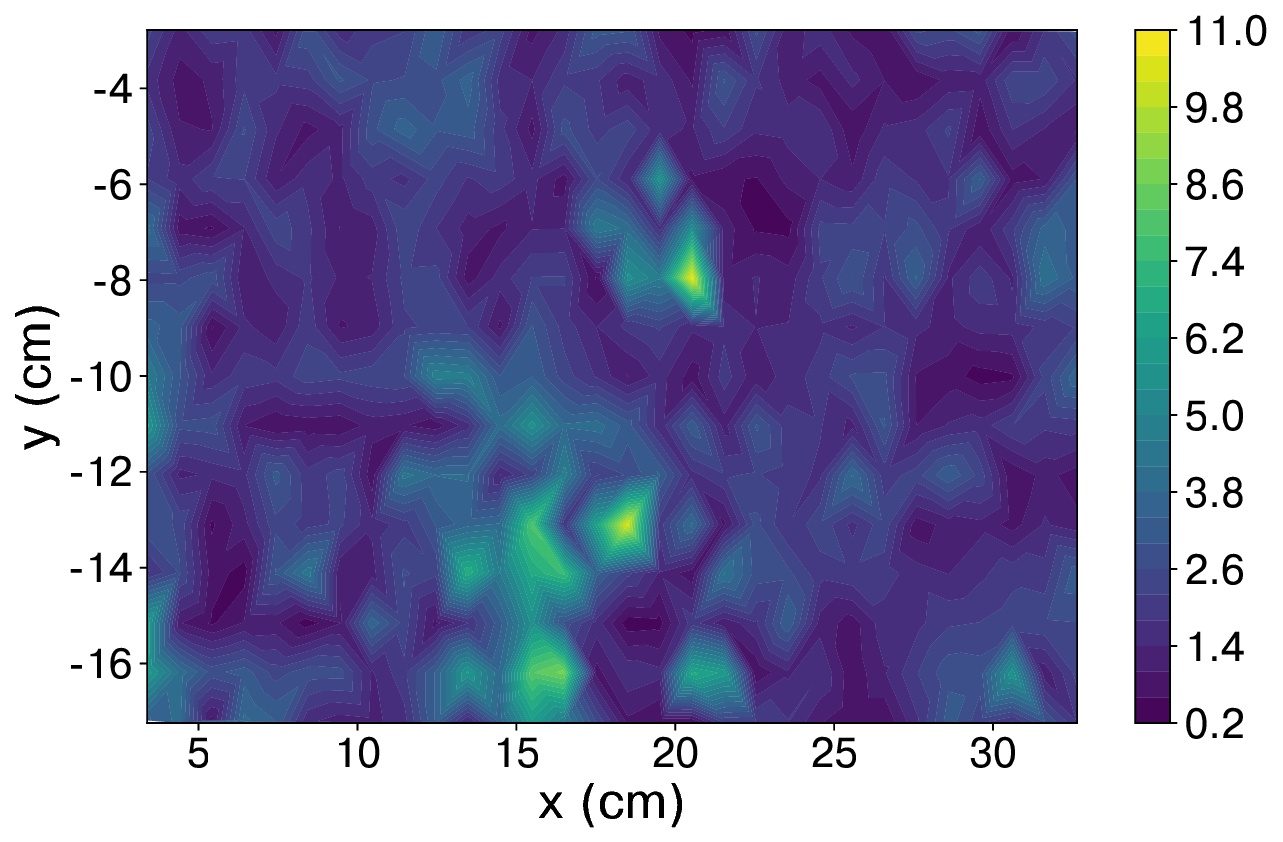}\\
		[-1ex]
		\caption{}
		\label{subfig:pos_err_rmse_csp}
	\end{subfigure}
	\hfill
	\begin{subfigure}[t]{0.48\textwidth}
		\raggedleft
		\includegraphics[width=\textwidth]{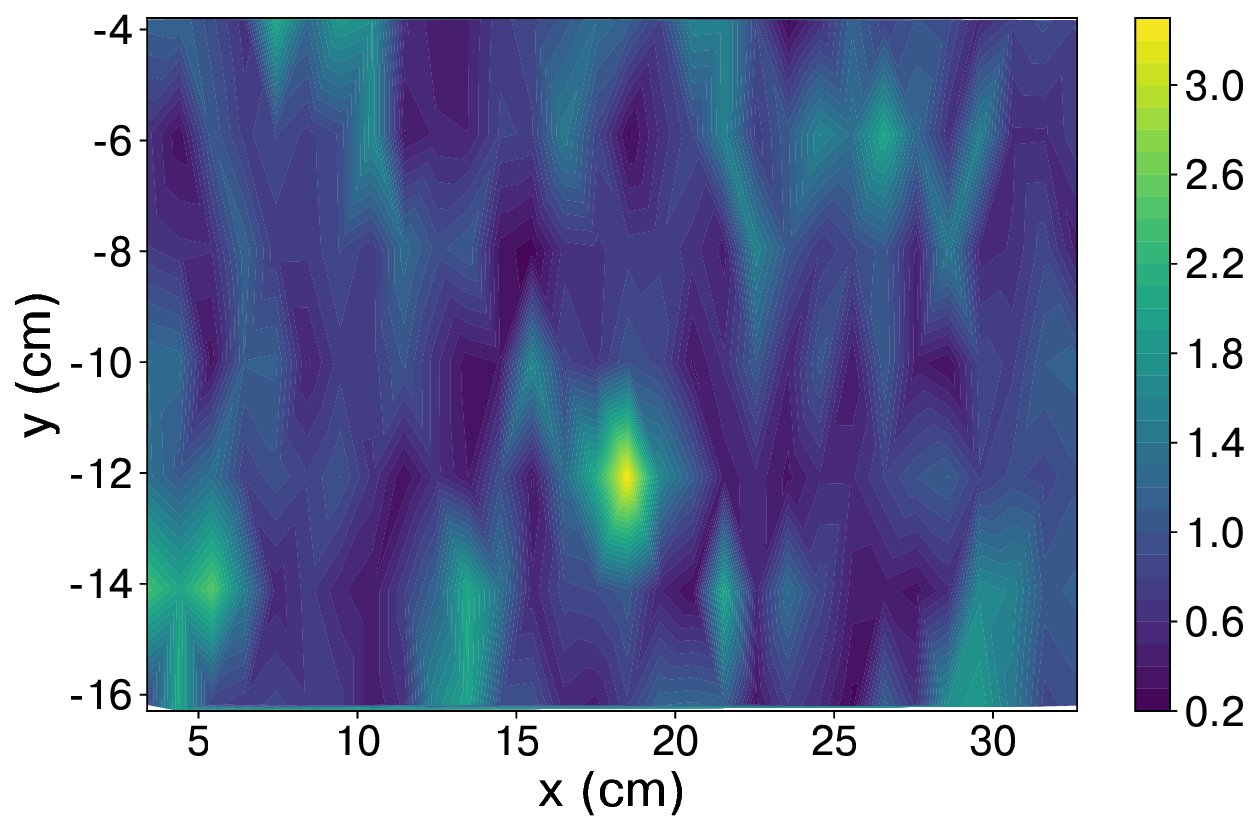}\\
		[-1ex]
		\caption{}
		\label{subfig:pos_err_rmse_ml}
	\end{subfigure}
	\caption{Comparison of the root mean square (RMSE) position error heatmap (colorbar unit is cm) when using (a) conventional signal processing (after range calibration), and (b) our proposed CNN Model (from Fig. \ref{fig:dnn_architecture}).} 
	\label{fig:pos_err_rmse_csp_vs_ml}
\end{figure*}

\subsection{Position Error Performance Comparison}
Fig. \ref{fig:pos_err_ML_vs_CSP_comparison} shows the marginal distribution of position error (marginalized across the entire test dataset) for (a) training/validation/test datasets (CNN-based approach), and (b) the test dataset (CSP-based approach). First, we observe that there is more than a $2\times$ improvement in the validation/test position error performance in the median and $90^\text{th}$ percentile value, when using our CNN-based approach. Furthermore, we observe that these values are well within the range resolution of the radar waveform ($\Delta r_{\rm res} = 3.075$ cm). On the other hand, we observe that while range calibration significantly improves the position error performance, the range-calibrated CSP-based approach achieves a $90^\text{th}$ percentile position error of $3.7$ cm, which is $\sim 20\%$ higher than $\Delta r_{\rm res}$. In alignment with our understanding of the physical limitations imposed by the radar waveform, the CSP-based method is unable to achieve sub-resolution accuracy. The key performance statistics are summarized in Table \ref{table:perf_comparison}.

Fig. \ref{fig:pos_err_rmse_csp_vs_ml} shows the heatmaps of RMSE position error for the CNN and CSP-based methods, for different touch regions on the display. These heatmaps are computed for the test dataset, visualized in Fig. \ref{fig:val_and_test_points}.  We observe that the CNN-based approach results in more than a $3\times$ improvement in the worst-case (maximum) RMSE, compared to that of the CSP-based approach. In general, there is more than a $2 \times$ improvement in RMSE position error for the CNN-based approach relative to that of CSP, when comparing the point-wise median and $90^\text{th}$ percentile RMSE, as shown in Table \ref{table:perf_comparison}. 

\begin{figure}[t]
	\centering
	\includegraphics[width=0.85\linewidth]{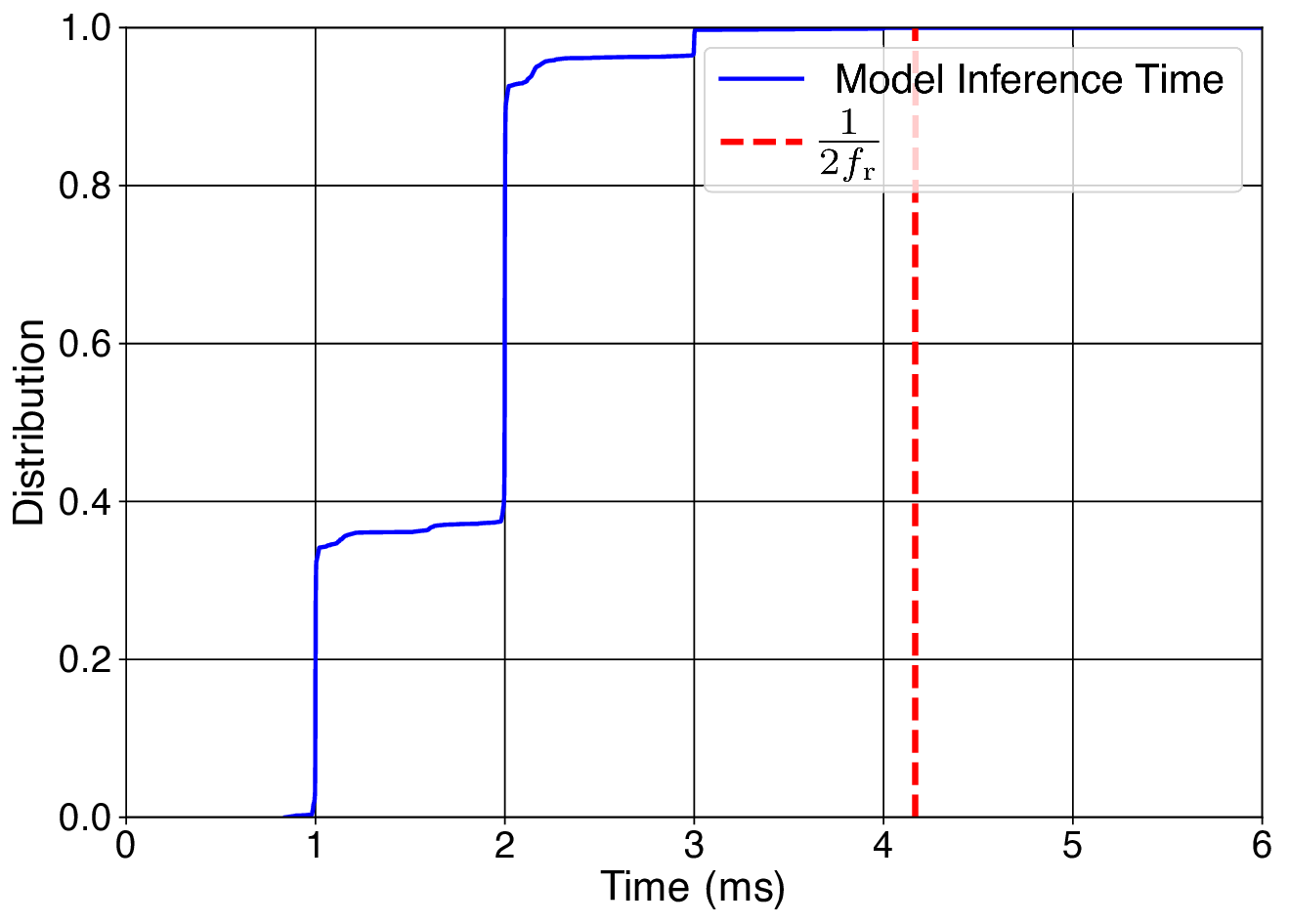}
	\caption{Distribution of the CNN inference time ($t_{\rm CNN, inf}$, shown in blue) versus the half radar frame repetition interval ($1/2f_{\rm r}$, shown in red).}
	\label{fig:model_inference_time}
\end{figure}

\subsection{Feasibility of Real-Time Implementation}
Inference execution time and model size are two important aspects of the CNN model that determine whether it is suitable for real-time implementation. Our model has $\sim 9 \times 10^4$ parameters, with a total size of $\sim 350$ KB. Thus, the memory required to store the model is quite small, and can be fitted on any standard system-on-chip (SoC). 

For integrating any ML-based algorithm into a real-time localization system, it is important that the inference time ($\rm t_{\rm CNN, inf}$) be smaller than the radar frame repetition interval ($1/f_{\rm r}$). To evaluate feasibility, we used a computer with an Intel i7-1185G7 processor, $16$ GB RAM, and no GPU. Fig. \ref{fig:model_inference_time} shows the distribution of $\rm t_{\rm CNN, inf}$ characterized on the test dataset. We observe that the median as well as the $90^\text{th}$ percentile inference time is $2$ ms, which is considerably smaller than the radar frame interval ($8.33$ ms, in our system). Thus, the small model size and inference time indicates that our proposed CNN-based approach is well-suited for real-time implementation on general purpose processor-based platforms. 
\vspace{-10pt}
\section{Conclusion}
In this paper, we proposed a mmWave FMCW radar-based touch localization system, wherein a deep neural network was trained to accurately localize a robot-mounted metal finger. We demonstrated that the CNN-based approach achieved sub-resolution position error, and significantly outperformed conventional signal processing-based algorithms. Finally, we discussed the feasibility of implementing our proposed approach in real-time. The small (a) CNN model size, and (b) inference time on general purpose computing platforms (relative to the radar frame interval), point towards a very strong feasibility for implementation on a real-time localization system. 

In this work, we have focused on accurate localization of robot-mounted targets. In general, extending this work to design localization systems for small targets such as accurate touch localization of human finger, and enabling handwriting on non touchscreen displays are worthwhile to enable low-cost technologies for human-screen interaction. 

\bibliographystyle{IEEEtran}
\bibliography{refs_ml_for_radar_touch}

\begin{thebibliography}{10}
\providecommand{\url}[1]{#1}
\csname url@samestyle\endcsname
\providecommand{\newblock}{\relax}
\providecommand{\bibinfo}[2]{#2}
\providecommand{\BIBentrySTDinterwordspacing}{\spaceskip=0pt\relax}
\providecommand{\BIBentryALTinterwordstretchfactor}{4}
\providecommand{\BIBentryALTinterwordspacing}{\spaceskip=\fontdimen2\font plus
\BIBentryALTinterwordstretchfactor\fontdimen3\font minus
  \fontdimen4\font\relax}
\providecommand{\BIBforeignlanguage}[2]{{%
\expandafter\ifx\csname l@#1\endcsname\relax
\typeout{** WARNING: IEEEtran.bst: No hyphenation pattern has been}%
\typeout{** loaded for the language `#1'. Using the pattern for}%
\typeout{** the default language instead.}%
\else
\language=\csname l@#1\endcsname
\fi
#2}}
\providecommand{\BIBdecl}{\relax}
\BIBdecl

\bibitem{Lin_CapTouch_Stylus_JDT_2013}
C.-L. Lin, C.-S. Li, Y.-M. Chang, T.-C. Lin, J.-F. Chen, and U.-C. Lin,
  ``{Pressure Sensitive Stylus and Algorithm for Touchscreen Panel},''
  \emph{Journal of Display Technology}, vol.~9, no.~1, pp. 17--23, 2013.

\bibitem{yun_strata_mobisys_2017}
S.~Yun, Y.-C. Chen, H.~Zheng, L.~Qiu, and W.~Mao, ``{Strata: Fine-Grained
  Acoustic-based Device-Free Tracking},'' in \emph{Proceedings of the 15th
  annual international conference on mobile systems, applications, and
  services}, 2017, pp. 15--28.

\bibitem{wu_fingerdraw_ACM_IMWUT_2020}
D.~Wu, R.~Gao, Y.~Zeng, J.~Liu, L.~Wang, T.~Gu, and D.~Zhang, ``{FingerDraw:
  Sub-Wavelength Level Finger Motion Tracking with WiFi Signals},''
  \emph{Proceedings of the ACM on Interactive, Mobile, Wearable and Ubiquitous
  Technologies}, vol.~4, no.~1, pp. 1--27, 2020.

\bibitem{wang_rf_idraw_acm_sigcomm_2014}
J.~Wang, D.~Vasisht, and D.~Katabi, ``{RF-IDraw: Virtual Touch Screen in the
  Air using RF Signals},'' \emph{ACM SIGCOMM Computer Communication Review},
  vol.~44, no.~4, pp. 235--246, 2014.

\bibitem{shangguan_Polarize_CoNEXT_2016}
L.~Shangguan and K.~Jamieson, ``{Leveraging Electromagnetic Polarization in a
  Two-Antenna Whiteboard in the Air},'' in \emph{Proceedings of the 12th
  International on Conference on emerging Networking EXperiments and
  Technologies}, 2016, pp. 443--456.

\bibitem{wei_mtrack_mobicom_2015}
T.~Wei and X.~Zhang, ``{mTrack: High-Precision Passive Tracking using
  Millimeter Wave Radios},'' in \emph{Proceedings of the 21st Annual
  International Conference on Mobile Computing and Networking}, 2015, pp.
  117--129.

\bibitem{xiao_milliback_acm_2019}
N.~Xiao, P.~Yang, X.-Y. Li, Y.~Zhang, Y.~Yan, and H.~Zhou, ``{MilliBack:
  Real-Time Plug-n-Play Millimeter Level Tracking using Wireless
  Backscattering},'' \emph{Proceedings of the ACM on Interactive, Mobile,
  Wearable and Ubiquitous Technologies}, vol.~3, no.~3, pp. 1--23, 2019.

\bibitem{cao_itracku_acm_2021}
Y.~Cao, A.~Dhekne, and M.~Ammar, ``{ITrackU: Tracking a Pen-like Instrument via
  UWB-IMU Fusion},'' in \emph{Proceedings of the 19th Annual International
  Conference on Mobile Systems, Applications, and Services}, 2021, pp.
  453--466.

\bibitem{Nermine_DL_AirWrite_IEEE_SJ_2022}
N.~Hendy, H.~M. Fayek, and A.~Al-Hourani, ``{Deep Learning Approaches for
  Air-Writing Using Single UWB Radar},'' \emph{IEEE Sensors Journal}, vol.~22,
  no.~12, pp. 11\,989--12\,001, 2022.

\bibitem{Faheem_in_air_write_IEEE_Acc_2020}
F.~Khan, S.~K. Leem, and S.~H. Cho, ``{In-Air Continuous Writing Using UWB
  Impulse Radar Sensors},'' \emph{IEEE Access}, vol.~8, pp. 99\,302--99\,311,
  2020.

\bibitem{buehrer2019handbook}
R.~Zekavat and R.~M. Buehrer, \emph{{Handbook of Position Location: Theory,
  Practice and Advances}}.\hskip 1em plus 0.5em minus 0.4em\relax Wiley-IEEE
  Press, 2019.

\bibitem{LeNet5_YLC_Proc_IEEE_1998}
Y.~Lecun, L.~Bottou, Y.~Bengio, and P.~Haffner, ``{Gradient-based Learning
  Applied to Document Recognition},'' \emph{Proceedings of the IEEE}, vol.~86,
  no.~11, pp. 2278--2324, 1998.

\bibitem{Shen_CAuth_IEEE_TKDE_2023}
Z.~Shen, S.~Li, X.~Zhao, and J.~Zou, ``{CT-Auth: Capacitive Touchscreen-Based
  Continuous Authentication on Smartphones},'' \emph{IEEE Transactions on
  Knowledge and Data Engineering}, pp. 1--16, 2023.

\bibitem{Yan_Tiberius_AESM_2013}
J.~Yan, C.~C. J.~M. Tiberius, G.~J.~M. Janssen, P.~J.~G. Teunissen, and
  G.~Bellusci, ``{Review of Range-based Positioning Algorithms},'' \emph{IEEE
  Aerospace and Electronic Systems Magazine}, vol.~28, no.~8, pp. 2--27, 2013.

\bibitem{Rao_Bey_Conv_Hull_TVT_2021}
R.~M. Rao, A.~V. Padaki, B.~L. Ng, Y.~Yang, M.-S. Kang, and V.~Marojevic,
  ``{ToA-Based Localization of Far-Away Targets: Equi-DOP Surfaces, Asymptotic
  Bounds, and Dimension Adaptation},'' \emph{IEEE Transactions on Vehicular
  Technology}, vol.~70, no.~10, pp. 11\,089--11\,094, 2021.

\bibitem{Rao_RNDOP_TVT_2024}
R.~M. Rao and D.-R. Emenonye, ``{Iterative RNDOP-Optimal Anchor Placement for
  Beyond Convex Hull ToA-Based Localization: Performance Bounds and Heuristic
  Algorithms},'' \emph{IEEE Transactions on Vehicular Technology}, vol.~73,
  no.~5, pp. 7287--7303, 2024.

\bibitem{Badriasl_3D_AoA_IEEE_TAES_2014}
L.~Badriasl and K.~Dogancay, ``{Three-Dimensional Target Motion Analysis using
  Azimuth/Elevation Angles},'' \emph{IEEE Transactions on Aerospace and
  Electronic Systems}, vol.~50, no.~4, pp. 3178--3194, 2014.

\bibitem{Santra_Hazra_DL_Radars_2020}
A.~Santra and S.~Hazra, \emph{{Deep Learning Applications of Short-Range
  Radars}}.\hskip 1em plus 0.5em minus 0.4em\relax Artech House, 2020.

\bibitem{Uysal_PC_FMCW_IEEE_TVT_2020}
F.~Uysal, ``{Phase-Coded FMCW Automotive Radar: System Design and Interference
  Mitigation},'' \emph{IEEE Transactions on Vehicular Technology}, vol.~69,
  no.~1, pp. 270--281, 2020.

\bibitem{FCC_Empows_60GHz_Rad}
\BIBentryALTinterwordspacing
{FCC}, ``{FCC Empowers Short-Range Radars in the 60 GHz Band},'' \emph{{Federal
  Communications Commission, Final Rule}}, July 2023. [Online]. Available:
  \url{https://www.govinfo.gov/content/pkg/FR-2023-07-24/pdf/2023-15367.pdf}
\BIBentrySTDinterwordspacing

\end{thebibliography}
\end{document}